%% file: main.tex
\documentclass{llncs}

\usepackage{graphicx}
\usepackage{tabularx}

\begin{document}

\title{Tracking Software Security Topics}

\author{Phong Minh Vu\inst{1} \and Tung Thanh Nguyen\inst{2}}

\institute{
Auburn University 
\email{lenniel@auburn.edu}\\
\and
Fulbright University Vietnam
\email{thanhtung.nguyen@fulbright.edu.vn}\\
}
\maketitle


\input{abstract}
\input{introduction}
\input{approach}
\input{evaluation}
\input{related}

\input{conclusion}

\bibliographystyle{splncs04}
\bibliography{references}

\end{document}

%% file: abstract.tex
\begin{abstract}
Software security incidents occur everyday and thousands of software security reports are announced each month. Thus, it is difficult for software security researchers, engineers, and other stakeholders to follow software security topics of their interests in real-time. In this paper, we propose, SOSK, a novel tool for this problem. SOSK allows a user to import a collection of software security reports. It pre-processes and extracts the most important keywords from the textual description of the reports. Based on the similarity of embedding vectors of keywords, SOSK can expand and/or refine a keyword set from a much smaller set of user-provided keywords. Thus, SOSK allows users to define any topic of their interests and retrieve security reports relevant to that topic effectively. Our preliminary evaluation shows that SOSK can expand keywords and retrieve reports relevant to user requests.   
\end{abstract}

%% file: introduction.tex
\section{Introduction}\label{sec:introduction}

Software security incidents occur everyday. While most are trivial, some high profile incidents  could lead to big financial losses or even human lives. To discover new software security vulnerabilities or enhance their software systems with latest security updates, software security researchers, engineers, and other stakeholders have to closely monitor software security topics of their interests. However, this is a difficult task because thousands of software security reports are announced each month. For example, in 2023, there are in total nearly 30,000 records newly added to CVE (Common Vulnerability and Exposures) - the most popular database of software security vulnerabilities\footnote{https://www.cvedetails.com/browse-by-date.php}.
The CVE website has a simple function for searching CVE reports. A user can provide one or several keywords to search for records containing those keywords in their textual description\footnote{https://cve.mitre.org/cve/search\_cve\_list.html}. However, it is still difficult for users, because they need to provide all relevant keywords to a specific topic of their interests.

To address this problem, in this paper, we propose  SOSK, a novel approach for extracting keywords and tracking security reports. SOSK allows a user to download a database of software security reports. It then pre-processes and extracts the most important keywords from the textual description of the reports. Based on the similarity of embedding vectors of keywords, SOSK can expand and/or refine a keyword set from a much smaller set of user-provided keywords. Thus, SOSK allows users to define any topic of their interests and retrieve security reports relevant to that topic effectively.

We conducted a preliminary evaluation on a dataset of 112,197 reports from the CVE database. After indexing, we used SOSK to define and expand keywords for three topics of common software vulnerabilities: \emph{SQL Injection}, \emph{Cross-site Scripting}, and \emph{Buffer Overflow}. We compared the trends of these three topics found by SOSK with a published report and found SOSK discovered the same trends. Then, we used SOSK to refine a topic of \emph{Mobile Devices} and retrieved the related CVE reports. The result shows that SOSK can expand keywords and retrieve reports relevant to our requests. We will perform more extensive evaluation on the performance and usefulness of our approach in the future.

In Section 2, we introduce our approach in detail, including techniques for indexing security reports and and tracking user-defined topics. Section 3 describes our evaluation settings and results. Section 4 presents related work and conclusions appear last.

%% file: approach.tex
\section{Approach}
In this section, we will discuss our approach in detail. SOSK is designed as a document management system with three main components: 1) a document store (DS); 2) a dictionary of keywords (KD); and 3) an index mapping keywords to documents (IX).

The \emph{document store} DS is a locally stored database containing the security reports that SOSK has imported and indexed. Each report has two main attributes: its created date and its textual content. Keywords are stored in a \emph{dictionary} KD. Each keyword has a textual value (like \textsf{sql} or \textsf{overflow}), a weighting score (like 0.95 or 0.73), and an embedding vector of 768 real values. To simplify the indexing and tracking steps, the textual values of keywords are converted to low-case.
The  \emph{index} IX stores for each keyword a hash map containing the ids of documents (security reports) containing that keyword and its positions in each document. Keywords and such positions are extracted from the documents' content while indexing. 

Because SOSK is designed for software security domain, in addition to common English words (nouns, verbs, adjectives), the keyword dictionary KD also contains domain-specific names and identifiers such as \textsf{windows}, \textsf{win32}, \textsf{sql}, \textsf{linux}, \textsf{ls} (a command in Linux), \textsf{http} (a network protocol), or  \textsf{svchost} (a process in Windows OS). We pre-populated this keyword dictionary with a dictionary of English words extracted from WordNet~\cite{miller1998wordnet} and a domain specific vocabulary for software documents collected from previous
studies~\cite{vu2015MARK,chen2014,deca}. Users can also import user predefined dictionaries to add more domain-specific vocabulary. 

SOSK has two main functions: \emph{indexing} and \emph{tracking}. When a user imports one or (typically) a collection of security reports into SOSK, it will perform indexing on imported documents to update its document store, keyword dictionary, and index. In tracking, SOSK allows a user to define a topic as a list of keywords, expand this list with more related keywords recommended by SOSK, and retrieve all documents relevant to such keywords. Let us discuss each function below.   

\subsection{Indexing}

SOSK performs indexing each document (security reports) by low-casing and tokenizing its textual content first. SOSK tries to correct each token using a common  for mapperly misspelled words and names provided by MARK~\cite{vu2015MARK}. For example, \textsf{watsapp} is corrected as \textsf{whatsapp} (the name of a popular mobile app for texting). Then, for each string token, if it exists in the keyword dictionary KD as a domain specific term (e.g., \textsf{windows} or \textsf{sql}), no stemming is applied. Otherwise, if it is a common English word, Snowball Stemmer is used for stemming. If the token does not exist in the dictionary, it is added as a new keyword. 

During this process, SOSK will update the index IX and the keyword dictionary KD for each keyword. The index IX will be updated to keep track the original positions of each keyword in the document under indexing. This is necessary in case users want to identify or highlight the analyzed text on the actual document later. The occurrence count of each keyword and co-occurrence count of each pair of keywords are also updated for calculating the keywords' scores and embedding vectors later.   

\subsubsection{Embedding vectors}

According to recent advances in natural language processing and language modeling, the semantic similarity of words can be measured by the distance of their embedding vectors. Thus, SOSK calculates for each keyword an embedding vector of 768 real numbers using Glove~\cite{glove}. Then, for two keyword $a,b$ with corresponding embedding vectors $v_a, v_b$, their similarity is define as $$\rho(a,b) = \frac {|v_a.v_b|}{||v_a||.||v_b||}$$




\subsubsection{Keyword scores} SOSK calculates for each keyword a score for later ranking. Currently, this score is its inverse document frequency (\emph{idf}). If the document store DS has in total $N$ documents and a keyword $a$ appears in $N_a$ documents, then its \emph{idf} score is $$d(a) = log(N/N_a)$$ 


\subsection{Tracking}

In SOSK, a topic is a set of keywords. A document contains one of some of those keywords will be considered to be relevant to that topic. For example, the set \{\textsf{buffer}, \textsf{memory}, \textsf{overflow}\} could be used to define the topic \emph{buffer overflow}. A security report contains such words will likely be written about vulnerabilities from buffer overflow.

SOSK calculates the relevance of a document to a topic based on the Vector Space Model. Assume a keyword $a$ occurs $n_{a,S}$ times in a document $S$ and $S$ has in total $n_S$ word occurrences. The \emph{term frequency} (tf) of keyword $a$ in document $S$ is defined as $$t(a, S) = n_{a,S}/n_S$$ The relevance of keyword $a$ to document $S$ is $t(a,S).d(a)$. Then, the relevance of a  topic $Q$ to a document $S$ is the total relevance from all its keywords $$\Gamma(Q,S) = \sum_{a \in Q} {t(a,S).d(a)}$$ 

That means, the more keywords are used to define a topic, the higher relevance is calculated for documents containing keywords of that topic. However, a user might not know all the keywords to define a topic of his interest. Therefore, a key function in SOSK is to recommend more keywords to expand a user-defined topic. For a topic $Q$, SOSK will calculate the list of recommended keywords: 
$$\chi(Q) = \{b \; | \; \exists a \in Q : \rho(a,b) > \theta \}$$

This list contains every keyword $b$ that is sufficiently similar (e.g. higher than a threshold $\theta$) to at least a keyword $a$ in $Q$. This list will be ranked descending by the keyword scores $d(b)$ and presented to the user. He can add some of those keywords into $Q$ and repeat the recommending step until satisfied.    

After the user defines a topic $Q$, SOSK uses the index IX to find every document $S$ containing at least a keyword in $Q$. Then, it calculates the relevance score $\Gamma(Q,S)$. The user can rank those documents by their relevance scores or by their created dates (e.g., to focus on more recent security reports first). 

%% file: evaluation.tex
\section{Evaluation}
\label{sec:eval}

In this section, we report a preliminary evaluation on a dataset of 112,197 security reports from the Common Vulnerability and Exposures (CVE) database, dated from 1999 to 2016. 

We first downloaded this dataset and imported it into SOSK for indexing. Then, we pre-loaded the embedding vectors for all applicable keywords (common English words and technical names) in SOSK with the pre-trained embedding vectors from the English Wikipedia dump. After that, the co-occurrence counts from SOSK indexing step were used to fine-tuning those embedding vectors. 

That task was needed to produce high quality embedding vectors for SOSK. Because the CVE dataset contains only very short and descriptive documents, it is not sufficient to learn high quality embedding vectors. Using pre-trained vectors and fine-tuned them using a domain-specific dataset reinforces the embedding vectors of common words that appear in both corpora (Wikipedia and CVE), while the technical or domain-specific keywords in the CVE dataset would have more significant embedding vectors.


\subsection{Studying previously reported security topics}

Neuhaus and Zimmermann published a study of trend analysis for security problems in the CVE dataset in 2009~\cite{neuhaus2010security}. However, as per technological advance, the security topics may have changed after that. Therefore we picked three major topics \emph{SQL Injection}, \emph{Cross-site Scripting} and \emph{Buffer Overflow} from their study and used our tool on more recent CVE reports to answer two questions: 1) How did their trends changed after 2009; and 2) What are the actual descriptions of the security problems for those topics?

\begin{figure*}[t]
\begin{tabular}{cc}
  \includegraphics[width=60mm]{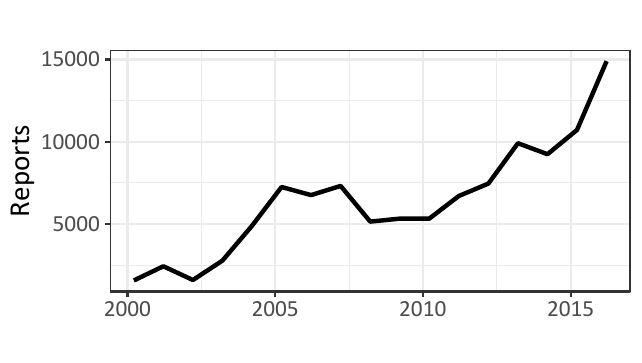} &   \includegraphics[width=60mm]{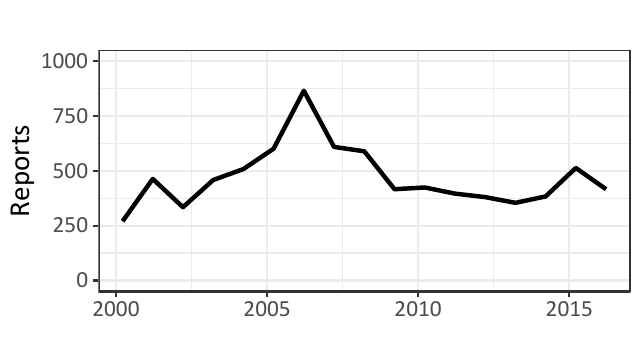} \\
(a) Total reports in CVE & (b) Buffer Overflow \\
 \includegraphics[width=60mm]{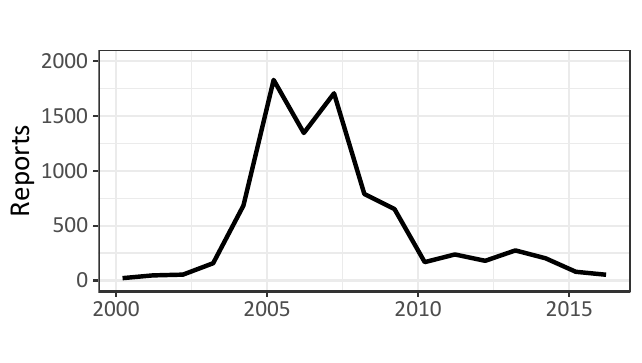} &   \includegraphics[width=60mm]{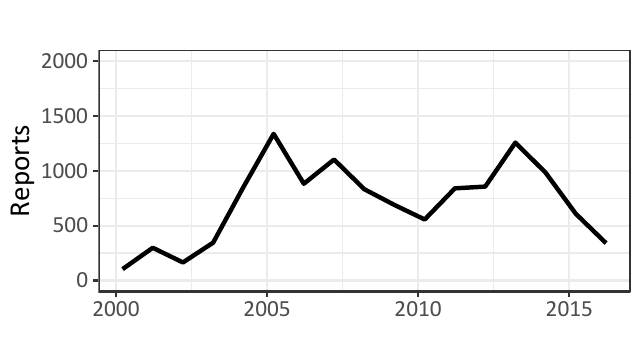} \\
(c) SQL Injection & (d) Cross-site Scripting \\
\end{tabular}
\caption{Trends analysis of three topics in CVE Dataset}
	\label{fig:cve}
\end{figure*}

The results are shown in Table \ref{tab:keywords}, Table \ref{tab:example.topicsexpansion}, and Figure \ref{fig:cve}. In Table 1, the  column \emph{Seeded Keywords} lists the keywords used by Neuhaus and Zimmermann in their study. We fed them into SOSK and SOSK recommended new keywords in the right column (with a similarity threshold of 0.9).

\begin{table*}[t]
\caption{Examples of Keywords Expansion}
\centering
\sf

    \begin{tabularx}{\textwidth}{|l| p{4cm}|X|}
\hline
    \textbf{Topic} & \textbf{Seeded Keywords} & \textbf{Expanded Keywords} \\
\hline
SQL Injection & sql, inject, vulnerability, php & sql, inject, vulnerability, php, \textit{vulnerable, code, arbitrary, command, injection, attacker, attack, remote}\\
\hline
Cross-site Scripting & vulnerability, script, cross, site, xss & vulnerability, script, cross, site, xss, \textit{website, server, remote, attacker, web, attack, directory, credential} \\
\hline
Buffer Overflow & overflow, buffer, stack, function & overflow, buffer, stack, function, \textit{component, method, memory, implement, implementation, heap, base}   \\
\hline
    \end{tabularx}%

\label{tab:keywords}
\end{table*}

\begin{figure}[th]
\centering
\includegraphics[width=80mm]{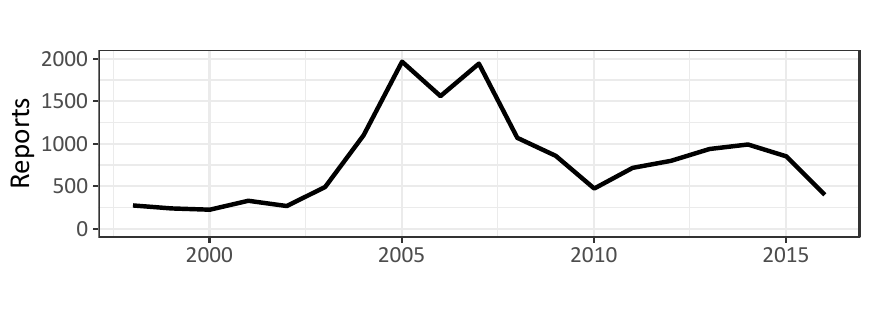}
\caption{Trends analysis of topic \emph{Mobile Devices} in CVE}
\label{fig:android}
\end{figure}

\subsubsection{How did security trends changed after 2009?}
As shown in Figure \ref{fig:cve}, even though we used a different approach, the shape of the trends from 2000 to 2009 reported by SOSK is similar to the findings reported in \cite{neuhaus2010security}. Then, after 2009 topic \emph{SQL Injection} had a sharp decline in the total number of reports, while the reports for \emph{Cross-site Scripting} fluctuated for the next few years. While the percentage of this type of attack was reduced dramatically during the those years, its actual number remained steady. Finally, after the gradual drop observed by Neuhaus and Zimmermann, \emph{Buffer Overflow} attacks kept declining slowly only to increased again during 2015 and 2016.
These numbers could indicate that \emph{SQL Injection} is not a popular breach around 2009 - 2016, but \emph{Cross-site Scripting} and \emph{Buffer Overflow} are still prevalent. 

\begin{table*}
\caption{Examples of Descriptions Extracted from Retrieved Reports}
\centering
\sf
\begin{tabularx}{\textwidth}{|l|X|}
\hline
\textbf{Topic} & \textbf{Examples of Topic Description} \\
\hline
SQL Injection & timecard cms allow remote attacker to execute arbitrary sql command  \\
        & execute arbitrary sql command via the execute query array  \\
        & myphpnuke allow remote attacker to execute arbitrary php code   \\
        & php in mybulletinboard allow remote attacker to execute arbitrary sql statement via the fidme    \\
        & sql injection vulnerability in aspwebalbum allow remote attacker to execute arbitrary sql statement via the username field    \\
\hline
Cross-site Scripting & site script vulnerability in the search function in the web management interface  \\
            & buymyscript lyric script allow remote attacker to inject arbitrary web script  \\
            & site script vulnerability in the calendar application \\
            & yahoo answer clone allow remote attacker to inject arbitrary web script  \\
            & site script attack by upload  \\
             
\hline
Buffer Overflow & buffer overflow in the urarlib get function  \\
            & multiple buffer overflow in the rtconfigload function  \\
            & heap buffer overflow in function pnmtoimage   \\
            & base buffer overflow in the imap server component    \\
            & stack buffer overflow in vshttpd   \\
\hline
\end{tabularx}
\label{tab:example.topicsexpansion}
\end{table*}

\subsubsection{What are the actual descriptions of the security reports?}
To answer the second research question, we printed out the descriptions matched keywords of the studying topics. Table \ref{tab:example.topicsexpansion} shows some examples. More can be viewed completely along with the artifacts of this paper\footnote{https://goo.gl/gNnk5i}. For \emph{SQL Injection}, we can see the the results listed the platforms or method used by attackers to inject arbitrary SQL code. For \emph{Cross-site Scripting} and \emph{Buffer Overflow}, we can see the vulnerabilities in different modules, software products, and components.

These results are informative enough to give users a general idea of what to be expected from each particular security problem, while saving the time needed to investigate directly in the CVE website. 

\subsection{Exploring new topic}

One topic not reported in the original study of Neuhaus and Zimmermann was security risk in \emph{mobile devices}. With CVE data at the time (from 1999 to 2009), the LDA method could not identify the rising of security reports related to mobile devices despite of a booming first year of iPhone being introduced to the world in 2017-2018. It is also not a pre-defined security topic at \textsf{cvedetails.com}.

However, in our keyword dictionary, the keyword \textsf{mobile} is among the highest ranked. To verify if this missing topic was truly important, we have applied our tool and the trend analysis method of MARK~\cite{vu2015MARK} (detecting abnormal spikes of Moving Average and Ratio of Difference to Standard Deviation) to see if it had a noticeable peak at some point in time, or somehow relevant at all. The results (Figure \ref{fig:android}) suggest that reports containing \textsf{mobile} and related keywords (\textsf{android}, \textsf{ios}, \textsf{iphone}, \textsf{samsung}) had a peak in 2005 and 2007, with 1965 and 1942 reports, respectively. After that, the yearly number of reports still remained high. This result has proven that this topic was overlooked by the LDA method and should have been a new important topic since 2005.

Further investigating reports returned by SOSK, we found that some of the most common vulnerabilities of mobile devices were \emph{"commonly available simple gps location"}, or \emph{"wi-fi spot configuration software"}, or \emph{"bypass intended permission restriction"}. More specific reports to the devices included: \emph{"samsung galaxy s4 through s7 device"}, \emph{"softback panasonic 3g handset"}, or \emph{"ip phone 1140e"}. Such information provides a coherent picture of which devices are vulnerable to which attacks, something specific to mobile devices for having various platforms.

%% file: related.tex
\section{Related work}

\label{sec:related}

Neuhaus and Zimmermann analyzed 39,393 vulnerability reports in the Common Vulnerability and Exposures (CVE) database from 1999 to 2009~\cite{neuhaus2010security}. They applied topic modeling (LDA) on descriptions of those CVE reports to find prevalent vulnerability types and new trends. In this work, we demonstrated that using a keyword-based approach can also find the same topics and trends.

MARK~\cite{vu2015MARK} is a keyword-based approach to discovering topics and trends on user reviews of mobile apps. In this paper, we reused its spell checker and domain-specific dictionary. However, due to domain  differences (reviews of mobile apps written by end-users vs security reports written by professionals), we had to enrich our keyword dictionary with more security-related terms and names. In addition, we also fine-tuned the pre-trained embedding vectors with textual contents of CVE reports to make it more relevant to the domain of software security. 

Yitagesu et al~\cite{tosem2023} developed unsupervised methods to label and extract important vulnerability concepts from the textual descriptions of software vulnerability reports. Their approach uses deep learning models (neural architecture that learns the Part-of-Speech (POS) of words and phrases and auto-encoder to encode syntactic similarities of paths in parse trees). The results are six \emph{fixed} types of vulnerability concepts. In contrast, our tool focuses on \emph{user-defined} topics which could be more flexible and suitable to users' interests. In addition, SOSK uses Glove~\cite{glove}, a lightweight approach for learning embedding vectors and thus, can index a large collection of data and run locally.

%% file: conclusion.tex
\section{Conclusions}
\label{sec:conclusion}
In this paper, we introduce SOSK, a tool for tracking software security reports by keywords. A user can collect a database of software security reports (typically text-based) and import it into SOSK. While importing, SOSK pre-processes and extracts the most important keywords from the textual description of the reports. Using pre-trained and fine-tuned embedding vectors of keywords, SOSK can compute their similarity and is able to expand and/or refine a keyword set from a much smaller set of user-provided keywords. Our preliminary evaluation shows that SOSK can enable users to define several topics of their interests and retrieve security reports relevant to that topic effectively.